\documentclass[notitlepage]{article}
\usepackage{graphicx}
\usepackage{amsmath}
\usepackage{amsfonts}
\usepackage{amssymb}
\begin{document}

\begin{center}
{\LARGE Hierarchic Theory of Condensed Matter:}

\smallskip

{\Large \ Long relaxation, macroscopic oscillations \smallskip}

{\Large and the effects of magnetic field}

\smallskip

\medskip

\textbf{Alex Kaivarainen}

\bigskip

\textbf{JBL, University of Turku, FIN-20520, Turku, Finland}

\medskip

{\large http://www.karelia.ru/\symbol{126}alexk}

{\large H2o@karelia.ru}

\medskip

\medskip
\end{center}

\begin{quotation}
\thinspace\thinspace\thinspace\thinspace\textbf{Materials, presented in this
original article are based on following publications:}

\smallskip

\textbf{[1]. A. Kaivarainen. Book: Hierarchic Concept of Matter and Field.
Water, biosystems and elementary particles. New York, NY, 1995, ISBN 0-9642557-0-7}

\textbf{[2]. \thinspace A. Kaivarainen. New Hierarchic Theory of Matter
General for Liquids and Solids: dynamics, thermodynamics and mesoscopic
structure of water and ice (see URL: http://www.karelia.ru/\symbol{126}alexk
\ [see New articles]).}

\textbf{[3].} \textbf{Hierarchic Concept of Condensed Matter and its
Interaction with Light: New Theories of Light Refraction, Brillouin
Scattering\ and M\"{o}ssbauer effect (http://www.karelia.ru/\symbol{126}alexk
\ \ [see New articles]). }

\textbf{[4]. A. Kaivarainen. Hierarchic Concept of Condensed Matter :
Interrelation between mesoscopic and macroscopic properties (see URL:
http://www.karelia.ru/\symbol{126}alexk \ \ [see New articles]). }

\bigskip

\textbf{See also set of previous articles at Los-Alamos archives: \ http://arXiv.org/find/physics/1/au:+Kaivarainen\_A/0/1/0/all/0/1}

\medskip

{\Large Contents:}

\smallskip

\textbf{Summary of new Hierarchic theory, general for liquids and solids.}

\textbf{1. Theoretical background for macroscopic oscillations in condensed matter}

\textbf{2. The hypothesis of [entropy - mass - time] interrelation}

\textbf{3. The entropy - information content of matter as a hierarchic system}

\textbf{4. Experimentally revealed macroscopic oscillations}

\textbf{5. Phenomena in water and aqueous systems, induced by magnetic field}

\textbf{\ \ Coherent radio-frequency oscillations in water, revealed by C. Smith}

\textbf{6 Influence of weak magnetic field on the properties of solid bodies}

\textbf{7 Possible mechanism of perturbations of nonmagnetic materials under
magnetic treatment}

\bigskip
\end{quotation}

\begin{center}
\textbf{\medskip\_\_\_\_\_\_\_\_\_\_\_\_\_\_\_\_\_\_\_\_\_\_\_\_\_\_\_\_\_\_\_\_\_\_\_\_\_\_\_\_\_\_\_\_\_\_\_\_\_\_\_\_\_\_\_\_\_\_\_\_\_\_\_\_\_\_\_\_\_\_\_\_\_\_\_\_}

{\Large Summary of new Hierarchic theory, general for liquids and solids}
\end{center}

{\large \smallskip}

\textbf{\ A basically new hierarchic quantitative theory, general for solids
and liquids, has been developed.}

\textbf{It is assumed, that unharmonic oscillations of particles in any
condensed matter lead to emergence of three-dimensional (3D) superposition of
standing de Broglie waves of molecules, electromagnetic and acoustic waves.
Consequently, any condensed matter could be considered as a gas of 3D standing
waves of corresponding nature. Our approach unifies and develops strongly the
Einstein's and Debye's models.}

\ \textbf{Collective excitations, like 3D standing de Broglie waves of
molecules, representing at certain conditions the mesoscopic molecular Bose
condensate, were analyzed, as a background of hierarchic model of condensed matter.}

\smallskip

\textbf{The most probable de Broglie wave (wave B) length is determined by the
ratio of Plank constant to the most probable impulse of molecules, or by ratio
of its most probable phase velocity to frequency. The waves B are related to
molecular translations (tr) and librations (lb).}

As the quantum dynamics of condensed matter does not follow in general case
the classical Maxwell-Boltzmann distribution, the real most probable de
Broglie wave length can exceed the classical thermal de Broglie wave length
and the distance between centers of molecules many times.

\textit{This makes possible the atomic and molecular Bose condensation in
solids and liquids at temperatures, below boiling point. It is one of the most
important results of new theory, which we have confirmed by computer
simulations on examples of water and ice.}

\smallskip

\textbf{Four strongly interrelated }new types of quasiparticles (collective
excitations) were introduced in our hierarchic model:

1.~\textit{Effectons (tr and lb)}, existing in "acoustic" (a) and "optic" (b)
states represent the coherent clusters in general case\textbf{; }

2.~\textit{Convertons}, corresponding to interconversions between \textit{tr
}and \textit{lb }types of the effectons (flickering clusters);

3.~\textit{Transitons} are the intermediate $\left[  a\rightleftharpoons
b\right]  $ transition states of the \textit{tr} and \textit{lb} effectons;

4.~\textit{Deformons} are the 3D superposition of IR electromagnetic or
acoustic waves, activated by \textit{transitons }and \textit{convertons. }\smallskip

\smallskip

\ \textbf{Primary effectons }(\textit{tr and lb) }are formed by 3D
superposition of the \textbf{most probable standing de Broglie waves }of the
oscillating ions, atoms or molecules. The volume of effectons (tr and lb) may
contain from less than one, to tens and even thousands of molecules. The first
condition means validity of \textbf{classical }approximation in description of
the subsystems of the effectons. The second one points to \textbf{quantum
properties} \textbf{of coherent clusters due to mesoscopic molecular Bose
condensation}\textit{. }

\ The liquids are semiclassical systems because their primary (tr) effectons
contain less than one molecule and primary (lb) effectons - more than one
molecule. \textit{The solids are quantum systems totally because both kind of
their primary effectons (tr and lb) are molecular Bose condensates.}%
\textbf{\ These consequences of our theory are confirmed by computer
calculations. }

\ The 1st order $\left[  gas\rightarrow\,liquid\right]  $ transition is
accompanied by strong decreasing of rotational (librational) degrees of
freedom due to emergence of primary (lb) effectons and $\left[
liquid\rightarrow\,solid\right]  $ transition - by decreasing of translational
degrees of freedom due to Bose-condensation of primary (tr) effectons.

\ \textbf{In the general case the effecton can be approximated by
parallelepiped with edges corresponding to de Broglie waves length in three
selected directions (1, 2, 3), related to the symmetry of the molecular
dynamics. In the case of isotropic molecular motion the effectons' shape may
be approximated by cube.}

\textbf{The edge-length of primary effectons (tr and lb) can be considered as
the ''parameter of order''.}

\smallskip

The in-phase oscillations of molecules in the effectons correspond to the
effecton's (a) - \textit{acoustic }state and the counterphase oscillations
correspond to their (b) - \textit{optic }state. States (a) and (b) of the
effectons differ in potential energy only, however, their kinetic energies,
impulses and spatial dimensions - are the same. The \textit{b}-state of the
effectons has a common feature with \textbf{Fr\"{o}lich's polar mode. }

\smallskip

\textbf{The }$(a\rightarrow b)$\textbf{\ or }$(b\rightarrow a)$%
\textbf{\ transition states of the primary effectons (tr and lb), defined
as\ primary transitons, are accompanied by a change in molecule polarizability
and dipole moment without density fluctuations. At this case they lead to
absorption or radiation of IR photons, respectively.}

\textbf{\ Superposition (interception) of three internal standing IR photons
of different directions (1,2,3) - forms primary electromagnetic deformons (tr
and lb).}

\ On the other hand, the [lb$\rightleftharpoons\,$tr] \textit{convertons }and
\textit{secondary transitons} are accompanied by the density fluctuations,
leading to \textit{absorption or radiation of phonons}.

\textit{Superposition resulting from interception} of standing phonons in
three directions (1,2,3), forms \textbf{secondary acoustic deformons (tr and
lb). }

\smallskip

\ \textit{Correlated collective excitations }of primary and secondary
effectons and deformons (tr and lb)\textbf{, }localized in the volume of
primary \textit{tr }and \textit{lb electromagnetic }deformons\textbf{, }lead
to origination of \textbf{macroeffectons, macrotransitons}\textit{\ }and
\textbf{macrodeformons }(tr and lb respectively)\textbf{. }

\ \textit{Correlated simultaneous excitations of \thinspace tr and lb}
\textit{macroeffectons }in the volume of superimposed \textit{tr }and
\textit{lb }electromagnetic deformons lead to origination of
\textbf{supereffectons. }

\ In turn, the coherent excitation of \textit{both: tr }and \textit{lb
macrodeformons and macroconvertons }in the same volume means creation of
\textbf{superdeformons.} Superdeformons are the biggest (cavitational)
fluctuations, leading to microbubbles in liquids and to local defects in solids.

\smallskip

\ \textbf{Total number of quasiparticles of condensed matter equal to 4!=24,
reflects all of possible combinations of the four basic ones [1-4], introduced
above. This set of collective excitations in the form of ''gas'' of 3D
standing waves of three types: de Broglie, acoustic and electromagnetic - is
shown to be able to explain virtually all the properties of all condensed matter.}

\ \textit{The important positive feature of our hierarchic model of matter is
that it does not need the semi-empiric intermolecular potentials for
calculations, which are unavoidable in existing theories of many body systems.
The potential energy of intermolecular interaction is involved indirectly in
dimensions and stability of quasiparticles, introduced in our model.}

{\large \ The main formulae of theory are the same for liquids and solids and
include following experimental parameters, which take into account their
different properties:}

$\left[  1\right]  $\textbf{- Positions of (tr) and (lb) bands in oscillatory spectra;}

$\left[  2\right]  $\textbf{- Sound velocity; }$\,$

$\left[  3\right]  $\textbf{- Density; }

$\left[  4\right]  $\textbf{- Refraction index (extrapolated to the infinitive
wave length of photon}$)$\textbf{.}

\textit{\ The knowledge of these four basic parameters at the same temperature
and pressure makes it possible using our computer program, to evaluate more
than 300 important characteristics of any condensed matter. Among them are
such as: total internal energy, kinetic and potential energies, heat-capacity
and thermal conductivity, surface tension, vapor pressure, viscosity,
coefficient of self-diffusion, osmotic pressure, solvent activity, etc. Most
of calculated parameters are hidden, i.e. inaccessible to direct experimental measurement.}

\ The new interpretation and evaluation of Brillouin light scattering and
M\"{o}ssbauer effect parameters may also be done on the basis of hierarchic
theory. Mesoscopic scenarios of turbulence, superconductivity and superfluity
are elaborated.

\ Some original aspects of water in organization and large-scale dynamics of
biosystems - such as proteins, DNA, microtubules, membranes and regulative
role of water in cytoplasm, cancer development, quantum neurodynamics, etc.
have been analyzed in the framework of Hierarchic theory.

\medskip

\textbf{Computerized verification of our Hierarchic theory of matter on
examples of water and ice is performed, using special computer program:
Comprehensive Analyzer of Matter Properties (CAMP, copyright, 1997,
Kaivarainen). The new optoacoustic device (CAMP), based on this program, with
possibilities much wider, than that of IR, Raman and Brillouin spectrometers,
has been proposed (see URL:\thinspace\thinspace
http://www.karelia.ru/\symbol{126}alexk).}

\smallskip

\textbf{This is the first theory able to predict all known experimental
temperature anomalies for water and ice. The conformity between theory and
experiment is very good even without any adjustable parameters. }

\textbf{The hierarchic concept creates a bridge between micro- and macro-
phenomena, dynamics and thermodynamics, liquids and solids in terms of quantum
physics. }

\bigskip

\begin{center}
=============================================================

\bigskip

{\large 1. Theoretical background for macroscopic oscillations in condensed matter}
\end{center}

\smallskip

\ \textbf{One of the consequences of our concept is of special interest. It is
the possibility for oscillation processes in solids and liquids. The law of
energy conservation is not violated thereupon because the energies of two
quasiparticle subsystems related to effectons and deformons, can change in
opposite phases. The total internal energy of matter keeps almost constant.}

The equilibrium shift between subsystems of condensed matter can be induced by
any external factor, i.e. pressure or field. The relaxation time, necessary
for system to restore its equilibrium, corresponding to minimum of potential
or free energy after switching off external factor can be termed ''memory'' of system.

\smallskip

\ \textbf{The energy redistribution between primary and secondary effecton and
deformon subsystems may have a periodical character, coupled with the
oscillation of the }$(a\Leftrightarrow b)$\textbf{\ equilibrium constant of
primary effectons }$(K_{a\Leftrightarrow b})$\textbf{\ and correlated
oscillations of primary electromagnetic deformons concentration if dissipation
processes are weak or reversible. According to our model (Table 1 of [1, 2])
the }$(a\rightarrow b)$\textbf{\ transition of primary effecton is related to
photon absorption, i.e. a decrease in primary electromagnetic deformon
concentration, while the }$(b\rightarrow a)$\textbf{\ transition on the
contrary, radiate photons. If, therefore, the }$\left[  a\Leftrightarrow
b\right]  $\textbf{\ and }$\left[  \bar{a}\Leftrightarrow\bar{b}\right]
$\textbf{\ equilibriums are shifted right ward, and equilibrium constants
}$K_{a\Leftrightarrow b}$\textbf{\ and }$\bar{K}_{a\Leftrightarrow b}%
$\textbf{\ decreases, then concentrations of primary and secondary deformons
}$(n_{d}$\textbf{\ and }$\bar{n}_{d})$\textbf{\ also decreases. If
}$K_{a\Leftrightarrow b}$\textbf{\ grows up, i.e. the concentration of primary
effectons in a-states increases, then }$n_{d}$\textbf{\ increases. We remind
that }$\left(  \mathbf{a}\right)  $\textbf{\ and }$\left(  \mathbf{b}\right)
$\textbf{\ states of the primary effectons correspond to the more and less
stable molecular clusters (see Introduction). In accordance with our model,
the strong interrelation exists between dynamic equilibrium of primary and
secondary effectons. Equilibrium of primary effectons is more sensitive to any
perturbations. However, the equilibrium shift of secondary effectons affect
the total internal energy, the entropy change and possible mass defect (see
below) stronger than that of primary effectons.}

\textbf{As we have shown (Fig. 28a,b of [1] and [3]), the scattering ability
of A-states is more than two times as high as that of B-states. Their
polarizability, refraction index and dielectric permeability are also higher.
It makes possible to register the oscillations in the condensed matter in
different ways.}

\textbf{In accordance with our theory the oscillation of refraction index must
induce the corresponding changes of viscosity and self-diffusion in condensed
matter (see Chapter 11 of [1] and [4]). The diffusion variations are possible,
for example, in solutions of macromolecules or other Brownian particles. In
such a way self-organization in space and time gradually may originate in
appropriate solvents, solutions, colloid systems and even in solid bodies.}

\textbf{The period and amplitude of these oscillations depend on the times of
relaxation processes which are related to the activation energy of equilibrium
shifts in the effectons, polyeffectons or coherent superclusters of primary
effectons subsystems.}

\textbf{The reorganizations in the subsystems of translational and librational
effectons, macro- and supereffectons, as well as chain-like polyeffectons,
whose stabilities and sizes differ from each other, must go on at different
rates. It should, therefore, be expected that in the experiment the presence
of several oscillation processes would be revealed. These processes are
interrelated but going with different periods and amplitudes. Concomitant
oscillations of self-diffusion rate also must be taken into account. In such a
way Prigogine's dissipative structures could be developed (Prigogine, 1984).
Instability in the degree of ordering in time and space is accompanied by the
slow oscillation of entropy of the whole macroscopic system.}

\textbf{\ The coherent extraterrestrial cosmic factors and gravitational
instabilities can induce long relaxation and oscillation processes in water
and other kind of condensed matter (Udaltsova, et. al., 1987).}

\medskip

\begin{center}
{\large 2. The hypothesis of [entropy - mass - time] interrelation}
\end{center}

\smallskip

The second law of thermodynamics for the closed system having the permanent
number of particles and the constant internal energy is:%

\begin{equation}
{\frac{dQ_{i}}{T}}=dS_{i}\ge0\tag{1}%
\end{equation}
where:$\;dQ_{i}$ appears to be due to the irreversible processes within the
system and is referred to as \textit{uncompensated heat }(Prigogine, 1980,
1984, Babloyantz, 1986).

This law means the only possibility of the processes, accompanied by the
increase of entropy (S) (related neither to chemical or nuclear reactions).

The statistical interpretation of entropy is expressed with the Boltzmann formula:%

\begin{equation}
S=\text{k}\cdot\text{lnP }\simeq\text{k}\cdot\text{lnW }\tag{2}%
\end{equation}

where: $k$ is the Boltzmann constant, P is the statistical weight, which is
proportional to the number of realizations (number of microstates) for the
given state of the macroscopic system (W).

Correspondingly the probability of one of this microstates is: $p_{i}=1/$W.

It follows from (2) that the second law of thermodynamics expresses the fact
that the system tends to the most probable state.%

\begin{equation}
dS=k\text{ }dlnW\ \ge0\tag{3}%
\end{equation}
or%

\begin{equation}
dQ_{i}=\text{ }TdS_{i}=kT\text{ }dlnW_{i}\tag{4}%
\end{equation}
Boltzmann has put forward the hypothesis that the irreversibility (i.e.
asymmetry) of time is determined with the irreversibility of processes
according to the second law of thermodynamics. Prigogine has modified and
developed this idea (Prigogine, 1984), introducing the notions of the
\textbf{internal time and microscopic operator of entropy}. The ''ARROW OF
TIME'' and irreversibility is a result of asymmetry of physically accessible
states after Prigogin. The microscopic mechanism of irreversibility is
discussed also in Chapter  8 of Part II of book [1].

It follows from the second law and formulae (1.12) from Part II of this book,
that the mass of a system of N similar particles with the mass $(M\simeq Nm)$
either does not change in the course of time, or decreases:%

\begin{equation}
{\frac{dt}{t}}=-{\frac{1}{2}}{\frac{dM}{M}}\ge0,\tag{5}%
\end{equation}

where: dM is the change in the mass of a macroscopic system related to
electromagnetic and acoustic radiation, and also to intermolecular interaction
induced small mass defect.

It follows from (5) that the positive pace of time in the system is related to
the decrease in its mass, while the negative pace of time - to the increase in
its mass. Hence, the coherent oscillation processes in the system (which are
not related to chemical reactions) are related to oscillation of mass [M]. It
is known that phonons and photons do not have the rest masses. Therefore, the
coherent periodical energy exchange between subsystems of effectons and
deformons (primary and secondary) must be accompanied by changes in the mass
of the effecton subsystem and total system.

The macroscopic fluctuations of mass values for solid bodies were registered
experimentally indeed (Kozyrev, 1958).

The oscillations of temperature and thermal conductivity also must accompany
the oscillation of secondary acoustic deformons concentration.

The total internal energy of the closed system can be represented as a sum of
contributions from the primary $(U^{a}$ and $U^{b})$, secondary $(\bar U^{a}$
and $\bar U^{b})$,\ macro $(U^{A},U^{B})$ and super $(U^{A^{*}},U^{B^{*}})$
effectons, convertons and contributions of corresponding deformons and
transitons (see formula 4.3).\ 

At $U_{\text{tot}}=$ $const$ \ the exchange of energies between the subsystems
of effectons and deformons can be approximately represented as:%

\begin{equation}
-d(mc_{x}^{2})N\approx d(h\overline{\bar{\nu}}_{d})N\tag{6}%
\end{equation}

where: $N$ \ is the number of particles in the system;

$\left[  c_{x}=(v_{gr}v_{ph})^{1/2}=const\right]  \;$ is the characteristic
wave B velocity in the system, which is equal to the product of the
generalized group $(v_{gr})$ and phase $(v_{ph})$ velocities of wave B of
particles; [$d(h\bar{\nu}_{d})N]$ is the change of energy contribution of
secondary deformons, which is much bigger than that of primary deformons due
to low concentration of latter; [$dm]$ is the defect of mass per particle in
the system as a result of energy exchange between subsystems of the effectons
and deformons.

\smallskip

Because phonons are carriers of the heat energy, then the uncompensated heat
$(dQ_{i})$ in eq.(1) could be easily related to the irreversible increment of
secondary acoustic deformons and convertons energy or to the uncertainty in
this energy:%

\begin{equation}
dQ_{i}=\text{ TdS }=Nd(h\overline{\overline{\nu}}_{d})_{tr,lb}\tag{7}%
\end{equation}

or%

\begin{equation}
dQ_{i}=\text{ TdS }=-d(mc_{x}^{2})N=-d(Mc_{x}^{2})_{tr,lb}\tag{8}%
\end{equation}%

\[
\text{where:\qquad}\overline{\bar\nu}_{d}=[(\bar\nu_{d})_{tr,lb}+(\bar\nu
_{d})_{ca,cb\text{,cMd}}]~+(\nu_{d})_{tr,lb}
\]

($\bar\nu_{d}$)$_{tr,lb}$ is the resulting frequency of secondary deformons
(tr and lb);

($\bar{\nu}_{d}$)$_{ca,cb\text{,cMd}}$ is the resulting frequency of
convertons and related deformons;

$M=Nm$ is the total mass of particles in the system.

($\nu_{d})_{tr,lb}$ are a frequencies of tr and lb IR photons.

\smallskip

\textbf{Let us enter the notion of absolute entropy (S) as a measure of the
uncompensated heat change }$(\Delta Q_{i}/T)$\textbf{, which could occur at
}$(\bar{a}\Leftrightarrow\bar{b})$\textbf{\ transitions of mean effectons and
convertons subsystems at the given temperature:}%

\begin{equation}
S=\int{\frac{dQ_{i}}{T}}\tag{9}%
\end{equation}
Putting (7) and (8) into (9) and multiplying numerator and denominator, to
Boltzmann constant (k) keeping in mind (5), we derive a new approximate
formula for absolute entropy:%

\begin{equation}
S={\frac{\Delta Q}{T}}={\frac{N\Delta\left(  3h\bar{\nu}_{d}\right)  }{T}%
}\approx3Nk\left(  {\frac{h\bar{\nu}^{a}}{kT}}\right)  \approx-{\frac{\Delta
Mc_{x}^{2}}{T}}={\frac{2M}{T}}{\frac{\Delta t}{t}}_{tr,lb}\tag{10}%
\end{equation}

where: $\bar\nu_{d}$ =$\left|  \bar\nu_{a}-\bar\nu_{b}\right|  \sim\bar\nu_{a}
$ is a frequency of secondary acoustic deformon; $\bar\nu_{a}$ is calculated
from formula (2.54).

Formula (10) relates the positive entropy value not only to heat radiation
from the system or the increase of the deformons contribution into the total
internal energy $(\Delta Q_{i}>0)$, determined the corresponding decrease of
its mass $(\Delta M<0)$, but also to positive time course in this system
$(\Delta t>0)$.

The interesting experiments done by Kosyrev (1958) confirm one of the
consequences of our theory, that the shift of equilibrium between subsystems
of the effectons and acoustic deformons to the latter ones should decrease the
mass of body. In his experiments with special balance it was shown that the
activation of phonons with sufficiently high energy by means of sound
generator - decreases the mass of solid body: $\left(  100-800\right)  \,g$.
The value of decreasing was proportional to the total mass of body. The
relative mass decreasing was estimated as:
\[
\Delta\mathbf{M}\mathbf{/}\mathbf{M}\mathbf{=}\mathbf{(}\mathbf{2}%
\mathbf{.}\mathbf{3}\mathbf{-}\mathbf{3}\mathbf{.}\mathbf{4}\mathbf{)}%
\cdot\mathbf{1}\mathbf{0}^{-5}%
\]

In liquids and solids, the concentration of secondary deformons can be lower
than the concentration of molecules. Therefore, in a general case N in the
formula (10) must be substituted by the number of secondary deformons in the system:%

\begin{equation}
(\overline{N}_{d})_{tr,lb}=V(\overline{n}_{d})_{tr,lb},\tag{11}%
\end{equation}
where V is the volume of the system;%

\[
(\bar n_{d})_{tr,lb}={\frac89}\pi\left(  {\frac{\bar\nu_{d}}{v_{\text{res}}}%
}\right)  _{tr,lb}^{3}
\]
is the concentration of secondary translational and librational effectons;%

\begin{equation}
\nu_{ph}^{res}=\left(  \bar{\nu}_{ph}^{1}\bar{\nu}_{ph}^{2}\bar{\nu}_{ph}%
^{3}\right)  ^{1/3}\tag{12}%
\end{equation}
is the resulting frequency of secondary deformons; $\bar{\nu}_{ph}^{1,2,3}$
are calculated from $(3.15);\;$ $v_{res}\;$is the resulting thermal phonons
velocity, which is equal to isotropic hypersonic velocity in liquids and to
transversal one $(v)$ in solids.

The total entropy of a condensed substance $(S_{\text{tot}})$ is approximately
a sum of secondary effecton contribution (\textit{tr and lb}) and contribution
\textit{of} convertons. Using (10) and (11), we obtain:%

\[
S_{\text{tot}}\approx S_{tr}+S_{lb}+S_{\text{con}}={\frac{V_{0}h}{T}}\left[
\begin{array}
[c]{c}%
\bar{n}_{d}^{tr}\left(
\begin{array}
[c]{c}%
3\bar{\nu}_{d}^{res}%
\end{array}
\right)  _{tr}+\bar{n}_{d}^{lb}\left(
\begin{array}
[c]{c}%
3\bar{\nu}_{d}^{res}%
\end{array}
\right)  _{lb}%
\end{array}
\right]  +
\]
\begin{equation}
+{\frac{V_{0}h}{T}}n_{\text{con}}3\left[
\begin{array}
[c]{c}%
(\nu_{ef}^{a})_{tr}-(\nu_{ef}^{a})_{lb}%
\end{array}
\right]  _{\text{con}}\tag{13}%
\end{equation}
where: $%
\begin{array}
[c]{l}%
\bar{n}_{d}^{tr}%
\end{array}
$ and $%
\begin{array}
[c]{l}%
\bar{n}_{d}^{lb}%
\end{array}
$ correspond to $(11);n_{\text{con}}$ is a concentration of convertons, equal
to that of primary librational effectons $(n_{lb})$;

($\overline{\nu}_{d}$)$_{tr,lb}^{res}$ are the resulting frequencies of
secondary deformons (tr and lb).

Knowing the positions of translational and librational bands in the
oscillatory spectra and sound velocity, one can estimate the entropy of matter
at each temperature using (13).

At constant pressure and volume the change of enthalpy (H) is equal to the
change of internal energy (U). Therefore, the formulae that we have obtained
for (U) and (S) allow to calculate also changes in free energy:%

\begin{equation}
\Delta G_{P,V}=\Delta U-T\Delta S\tag{14}%
\end{equation}
The qualitative correctness of the formula obtained for entropy is obvious
from (10) in the form:%

\begin{equation}
S_{tr,lb}\approx Nk(3h\overline{\nu}^{a}/kT)_{tr,lb}+Nk(3h\Delta\nu
^{a}/kT)_{\text{acon}}\simeq Nk(3h\nu^{a}/kT)_{tr,lb}\tag{15}%
\end{equation}
It follows from (2.54) and (15) that:

a) at $\,T\rightarrow0:\;h\bar{\nu}^{a}/kT\rightarrow0$ \ and\ \ $S\rightarrow0;$

b) at decreasing $\nu_{p}$ in the melting point at the growth of
$\;T$\ \thinspace both\thinspace\ $h\bar{\nu}_{a}/kT$ \ and\ $S$\ grow up;

c) if the mixing of liquids and gases leads to the weakening of pair
interaction in effectons or clusters, then $\,\nu_{p}=(E_{b}-E_{a})/h\,$
decreases and $S$ increases.

Equalizing (2) and right part of (15) we have in the framework of our approximations:%

\begin{equation}
S=k\ln P\approx k\left(  N_{0}\frac{h\overline{\nu}_{a}}{kT}\right)
=R\frac{h\overline{\nu}_{a}}{kT}\tag{16}%
\end{equation}
from (16) we can derive an approximate formula for statistical weight:%

\begin{equation}
P\approx\exp\left(  N_{0}{\frac{h\bar{\nu}_{a}}{kT}}\right)  _{tr,lb}\tag{17}%
\end{equation}
According to (16) the law of entropy growth \textit{means the striving of the
real quantum properties of the substance to ideal properties. }It means the
tending of primary effectons energy in the (a) state $(h\nu_{a})$ to the
thermal equilibrium value (kT):%

\begin{equation}
\Delta S>0\text{, \ \ if\ \ \ }[h\overline{\nu}^{a}\rightarrow kT]\tag{18}%
\end{equation}
\textbf{Competition between the discrete quantum energy distribution and its
tendency to kT may be a reason for instability of different parameters of
condensed matter. This may lead to origination of macroscopic oscillations in
the system interrelated with entropy, temperature and mass (eq.10). Such
oscillations could be considered as a kind of self-organization process due to
feedback links in a hierarchic system of interrelated quasiparticles of matter.}

\medskip

\begin{center}
{\large 3. The entropy - information content of matter as a hierarchic system}
\end{center}

\smallskip

\textbf{The statistical weigh for macrosystem (P), equal to number of
microstates (W), corresponding to given macrostate, necessary for entropy
calculation using (2) could be presented as:}%

\begin{equation}
W={\frac{N!}{N_{1}!\cdot N_{2}!\cdot\ldots\cdot N_{q}!}}\tag{19}%
\end{equation}
\textbf{where:}%

\begin{equation}
N=N_{1}+N_{2}+...N_{q}\tag{20}%
\end{equation}
\textbf{is the total number of molecules in macrosystem;}

$N_{i}$\textbf{\ is the number of molecules in the i-th state;}

$q$\textbf{\ is the number of independent states of all quasiparticles in macrosystem.}

\smallskip

We can subdivide macroscopic volume of $1cm^{3}$ into 24 types of
quasiparticles in accordance with our hierarchic model (see Table 1 of [1, 2]).

In turn, each type of the effectons (primary, secondary, macro- and
supereffectons) is subdivided on two states: ground (a,A) and excited (b,B)
states. Taking into account two ways of the effectons origination - due to
thermal translations (tr) and librations (lb), excitations, related to
$[lb/tr]$ convertons, macro- and super deformons, the total number of
\textbf{independent} states is 24 also. It is equal to number of independent
relative probabilities of excitations, composing partition function Z (see
eq.4.2 of [1, 2]). Consequently, in eqs.(4.19 and 4.20) we have:%

\[
q=24
\]
The \textbf{number }of molecules, in the unit of volume of condensed matter
(1cm$^{3})$, participating in each of 24 excitation states (i) can be
calculated as:%

\begin{equation}
N_{i}={\frac{(v)_{i}}{V_{0}/N_{0}}}\cdot n_{i}\cdot{\frac{P_{i}}{Z}}%
={\ \frac{N_{0}}{V_{0}}}{\frac{P_{i}}{Z}}\tag{21}%
\end{equation}
where: $(v)_{i}=1/n_{i}$ is the volume of (i) quasiparticle, equal to
reciprocal value of its concentration $(n_{i});\;N_{0}$ and $V_{0}$ are
Avogadro number and molar volume, correspondingly;\ Z is partition function
and $P_{i}$ are relative probabilities of independent excitations in
composition of $Z\;(eq.4.2)$.

The total number of molecules of (i)-type of excitation in any big volume of
matter $(V_{\text{Mac}})$ is equal to%

\begin{equation}
N_{\text{Mac}}^{i}=N_{i}V_{\text{Mac}}=V_{\text{Mac}}{\frac{N_{0}}{V_{0}}%
}{\frac{P_{i}}{Z}}\tag{21a}%
\end{equation}
Putting (20) into (18) and (19), we can calculate the statistical weight and
entropy from eq.(2).

For large values of N$_{i}$ it is convenient to use a Stirling formula:%

\begin{equation}
N_{i}=(2\pi N)^{1/2}(N/e)^{N}\cdot\exp(\Theta/12N)\sim(2\pi N)^{1/2}%
(N/\Theta)^{N}\tag{21b}%
\end{equation}
\medskip Using this formula and (20), one can obtain the following expression
for entropy:%

\begin{equation}
S=k\cdot\ln W=-k\cdot\sum_{i}^{q}(N_{i}+{\frac{1}{2}})\ln N_{i}+\text{ const
}=S_{1}+S_{2}+...S_{i}\tag{22}%
\end{equation}
From this eq. we can see that the temperature increasing or [solid
$\rightarrow$ liquid] phase transition will lead to the entropy elevation:%

\begin{equation}
\Delta S=S_{L}-S_{S}=k\cdot\ln(W_{L}/W_{S})>0\tag{23}%
\end{equation}
\medskip It follows from (22, 20) and (19) that under conditions when
$(P_{i})$ and $N_{i}$ undergoes oscillations it can lead to oscillations of
contributions of different types of quasiparticles to the entropy of system
and even to oscillations of total entropy of system as an additive parameter.
The coherent oscillations of $P_{i}$ and $N_{i}$ can be induced by different
external fields: acoustic, electromagnetic and gravitational. Macroscopic
autooscillations may arise spontaneously also in the sensitive and highly
cooperative systems.

Experimental evidence for such phenomena will be discussed in the next section.

The notions of probability of given microstate $(p_{i}= 1/W)$, entropy
$(S_{i})$ and information $(I_{i})$ are strongly interrelated. The smaller the
probability the greater is information (Nicolis 1986):%

\begin{equation}
I_{i}=\lg_{2}{\frac{1}{p}}_{i}=-\lg_{2}p_{i}=\lg_{2}W_{i}\tag{24}%
\end{equation}
where $p_{i}$ is defined from the Boltzmann distribution as:%

\begin{equation}
p_{i}={\frac{\exp(-E_{i}/kT)}{\sum_{m=0}^{\infty}\exp(-n_{m}h\nu_{i}/kT)}%
}\tag{25}%
\end{equation}
where\ n$_{m}$ is quantum number; h is the Plank constant; $E_{i}=h\nu_{i}$ is
the energy of (i)-state.

There is strict relation between the entropy and information, leading from
comparison of (24) and (2):%

\begin{equation}
S_{i}=(k_{B}\ln2)I_{i}=2.3\cdot10^{-24}I_{i}\tag{26}%
\end{equation}
The information entropy is given as expectation of the information in the
system (Nicolis,1986; Haken, 1988).%

\begin{equation}
<I>=\Sigma P_{i}\lg_{2}(1/p_{i})=-\Sigma p_{i}\lg_{2}(p_{i})\tag{27}%
\end{equation}
From (26) and (22) we can see that variation of probability $p_{i}$ and/or
$N_{i}$ in (20) will lead to changes of entropy and information,
characterizing the matter as a hierarchical system.

The \textbf{reduced information }(entropy), characterizing its
\textbf{quality}, related to selected collective excitation of any type of
condensed matter, we introduce here as a product of corresponding component of
information $[I_{i}]$ to the number of molecules (atoms) with similar dynamic
properties in composition of this excitation:%

\begin{equation}
q_{i}=(v_{i}/v_{m})=N_{0}/(V_{0}n_{i})\tag{27a}%
\end{equation}
where: $v_{i}=1/n_{i}$ is the volume of quasiparticle, reversible to its
concentration $(n_{i});\;v_{m}=V_{0}/N_{0}$ is the volume, occupied by one molecule.

The product of (27) and (27a), i.e. the \textbf{reduced information }gives the
quantitative characteristic not only about quantity but also about the quality
of the information:%

\begin{equation}
(Iq)_{i}=p_{i}\lg{}_{2}(1/p_{i})\cdot N_{0}/(V_{0}n_{i})\tag{27b}%
\end{equation}
\textbf{This new formula could be considered as a useful modification of known
Shennon equation.}

\medskip

\begin{center}
{\large 4. Experimentally revealed macroscopic oscillations}
\end{center}

\smallskip

A series of experiments was conducted in our laboratory to study oscillations
in the buffer (pH 7.3) containing 0.15 M NaCl as a control system and
immunoglobulin G solutions in this buffer at the following concentrations:
$3\cdot10^{-3};\;6\cdot10^{-3};\;1.2\cdot10^{-2}$ and $2.4\cdot10^{-2}mg/ml$.

The turbidity $(D^{*})$ of water and the solutions were measured every 10
seconds with the spectrophotometer at $\lambda=350nm$. Data were obtained
automatically with the time constant 5 s during 40 minutes. The number of
$D^{*}$ values in every series was usually equal to 256. The total number of
the fulfilled series was more than 30.

The time series of D$^{*}$ were processed by the software for time series
analysis. The time trend was thus subtracted and the autocovariance function
and the spectral density were calculated.

The empty quartz cuvette with the optical path about 1 cm were used as a basic control.

Only the optical density of water and water dissolved substances, which really
exceeded background optical density in the control series were taken into
account. It is shown that the noise of the photoelectronic multiplier does not
contribute markedly to dispersion of D$^{*}$.

The measurements were made at temperatures of $17,28,32$, and $37^{0}$. The
period of the trustworthily registered oscillation processes related to
changes in $D^{*}$, had 2 to 4 discrete values over the range of $\left(
30-600\right)  \,s$ under our conditions. It does not exclude the fact that
the autooscillations of longer or shorter periods exist. For example, in
distilled water at $32^{0}C$ the oscillations of the scattering ability are
characterized by periods of 30, 120 and 600 s and the spectral density
amplitudes 14, 38 and 78 (in relative units), respectively. With an increase
in the oscillation period their amplitude also increases. At $28^{0}C$ the
periods of the values 30, 41 and 92s see have the corresponding normalized
amplitudes 14.7, 10.6 and 12.0.

Autooscillations in the buffer solution at $28^{0}C$ in a 1 cm wide cuvette
with the optical way length 1 cm (i.e. square section) are characterized with
periods: $34,\,52,\,110$ and 240 s and the amplitudes: $24,\,33,\,27$ and 33
relative units. In the cuvette with a smaller (0.5 cm) or larger (5 cm)
optical wavelength at the same width (1 cm) the periods of oscillations in the
buffer change insignificantly. However, amplitudes decreased by 50\% in the 5
cm cuvette and by 10-20\% in the 0.5 cm-cuvette. This points to the role of
geometry of space where oscillations occur, and to the existence of the finite
correlation radius of the synchronous processes in the volume. But this radius
is macroscopic and comparable with the size of the cuvette.

The dependence of the autooscillations amplitude on the concentration of the
protein - immunoglobulin G has a sharp maximum at the concentration of
$1.2\cdot10^{-2}\,mg/ml$. There is a background for considering it to be a
manifestation of the hydrodynamic Bjorkness forces between the pulsing
macromolecules (K\"{a}iv\"{a}r\"{a}inen, 1987).

Oscillations in water and water solutions with nearly the same periods have
been registered by the light-scattering method by Chernikov (1985).

Chernikov (1990d) has studied the dependence of light scattering fluctuations
on temperature , mechanic perturbation and magnetic field in water and water
hemoglobin and DNA solution. It has been shown that an increase in temperature
results in the decline of long-term oscillation amplitude and in the increase
of short-time fluctuation amplitude. Mechanical mixing removes long- term
fluctuations and over 10 hours are spent for their recovery. Regular
fluctuations (oscillations) appear when the constant magnetic field above
$240A/m$ is applied; the fluctuations are retained for many hours after
removing the field. The period of long-term oscillations has the order of $10$
minutes. It has been assumed that the maintenance of long-range correlation of
molecular rotation-translation fluctuation underlies the mechanism of
long-term light scattering fluctuations.

It has been shown (Chernikov, 1990b) that a pulsed magnetic field (MF), like
constant MF, gives rise to light scattering oscillations in water and other
liquids containing H atoms: glycerin, xylol, ethanol, a mixture of unsaturated
lipids. All this liquids also have a distinct response to the constant MF.
''Spontaneous'' and MF-induced fluctuations are shown to be associated with
the isotropic component of scattering. These phenomena do not occur in the
nonproton liquid (carbon tetrachloride) and are present to a certain extent in
chloroform (containing one hydrogen atom in its molecule). The facts obtained
indicate an important role of hydrogen atoms and cooperative system of
hydrogen bonds in ''spontaneous'' and induced by external perturbations
macroscopic oscillations.

The understanding of such phenomena can provide a physical basis for of
self-organization (Prigogine, 1980, 1984, Babloyantz, 1986), the biological
system evolution (Shnol, 1979, Udaltsova et al., 1987), and chemical processes
oscillations (Field and Burger, 1988).

It is quite probable that macroscopic oscillation processes in biological
liquids, e.g. blood and liquor, caused by the properties of water are involved
in animal and human physiological processes.

We have registered the oscillations of water activity in the protein-cell
system by means of light microscopy using the apparatus "Morphoquant", through
the change of the erythrocyte sizes, the erythrocytes being ATP-exhausted and
fulfilling a role of the passive osmotic units. The revealed oscillations have
a few minute-order periods.

Preliminary data obtained from the analysis of oscillation processes in the
human cerebrospinal liquor indicate their dependence on some pathology.
Perhaps, the autooscillations spectrum of the liquor can serve as a sensitive
test for the physiological status of the organism. The liquor is an
electrolyte and its autooscillations can be modulated with the electromagnetic
activity of the brain.

\textbf{We suggested that the activity of the central nervous system and the
biological rhythms of the organism are dependent with the oscillation
processes in the liquor. If it is the case, then the directed influence on
these autooscillation processes, for example, by means of magnetic field makes
it possible to regulate the state of the organism and its separate organs.
Some of reflexotherapeutic effects can be caused by correction of biorhythms.}

\smallskip

During my stay in laboratory of Dr. G.Salvetty in the Institute of Atomic and
Molecular Physics in Pisa (Italy) in 1992, the oscillations of heat capacity
$[C_{p}]$ in 0.1 M phosphate buffer (pH7) and in 1\% solution of lysozyme in
the same buffer at $20^{0}C$ were revealed. The sensitive adiabatic
differential microcalorimeter was used for this aim. The biggest relative
amplitude changing: $[\Delta C_{p}]/[C_{p}]{\ }\sim(0.5\pm0.02)\%$ occurs with
period of about 24 hours, i.e. corresponds to circadian rhythm.

\textbf{Such oscillations could be stimulated by the variation of magnetic and
gravitational conditions of the Earth during this period.}

\medskip

\begin{center}
{\large 5. Phenomena in water and aqueous systems, induced by magnetic field}
\end{center}

\smallskip

In the works of (Semikhina and Kiselev, 1988, Kiselev et al., 1988, Berezin et
al., 1988) the influence of the weak magnetic field was revealed on the
dielectric losses, the changes of dissociation constant, density, refraction
index, light scattering and electroconductivity, the coefficient of heat
transition, the depth of super-cooling for distilled water and for ice also.
This field used as a modulator a geomagnetic action.

The absorption and the fluorescence of the dye (rhodamine 6G) and protein in
solutions also changed under the action of weak fields on water. The latter
circumstance reflects feedback links in the guest-host, or solute -solvent system.

The influence of constant and variable magnetic fields on water and ice in the
frequency range $10^{4}-10^{8}Hz$ was studied. The maximum sensitivity to
field action was observed at the frequency $\nu_{\max}=10^{5}Hz$. In
accordance with our calculations, this frequency corresponds to frequency of
superdeformons excitations in water (see Fig. 48d of [1] and article [2]).

A few of physical parameters changed after the long (nearly 6 hour) influence
of the variable fields (\~H), modulating the geomagnetic field of the tension
$[H=H_{\text{geo}}]$ with the frequency (\textit{f}) in the range of
$(1-10)\cdot10^{2}Hz\;($Semikhina and Kiselev, 1988, Kiselev et al., 1988):%

\begin{equation}
H=H\cos2\pi ft\tag{28}%
\end{equation}
In the range of modulating magnetic field (H) tension from $0.08$ $A/m$ to
$212A/m$ the \textbf{eight maxima of dielectric losses tangent }in the above
mentioned (\textit{f}) range were observed. Dissociation constant decreases
more than other parameters (by 6 times) after the incubation of ice and water
in magnetic field. The relaxation time (''memory'') of the changes, induced in
water by fields was in the interval from 0.5 to 8 hours.

The authors interpret the experimental data obtained as the influence of
magnetic field on the probability of proton transfer along the net of hydrogen
bonds in water and ice, which lead to the deformation of this net.

The \textit{equilibrium constant }for the reaction of dissociation:
\[
H_{2}O\Leftrightarrow OH^{-}+H^{+}
\]
in ice is less by almost six orders $(\simeq10^{6})$ than that for water. On
the other hand the values of the \textit{field}- \textit{induced effects in
ice are several times more than in water}, and the time for reaching them in
ice is less. So, the above interpretation is doubtful.

\textbf{In the framework of our concept all the aforementioned phenomena could
be explained by the shift of the }$(a\Leftrightarrow b)$\textbf{\ equilibrium
of primary translational and librational effectons to the left.} In turn, this
shift stimulates polyeffectons or coherent superclusters growth, under the
influence of magnetic fields. Therefore, parameters such as the refraction
index, dielectric permeability and light scattering have to enhance
symbatically, while the $H_{2}O$ dissociation constant depending on the
probability of superdeformons must decrease. The latter correlate with
declined electric conductance.

\textbf{As far, the magnetic moments of molecules within the coherent
superclusters or polyeffectons formed by primary librational effectons are
additive, then the values of changes induced by magnetic field must be
proportional to polyeffecton sizes. These sizes are markedly higher in ice
than in water and decrease with increasing temperature.}

Inasmuch the effectons and polyeffectons interact with each other by means of
phonons (i.e. the subsystem of secondary deformons), and the velocity of
phonons is higher in ice than in water, then the saturation of all concomitant
effects and achievement of new equilibrium state in ice is faster than in water.

The frequencies of geomagnetic field modulation, at which changes in the
properties of water and ice have maxima can correspond to the
eigen-frequencies of the $\left[  a\Leftrightarrow b\right]  $ equilibrium
constant of primary effectons oscillations, determined by [assembly
$\Leftrightarrow$ disassembly] equilibrium oscillations for coherent super
clusters or polyeffectons.

The presence of dissolved molecules (ions, proteins) in water or ice can
influence on the initial $[a\Leftrightarrow b]$ equilibrium dimensions of
polyeffectons and,consequently the interaction of solution with outer field.

Narrowing of $^{1}$H-NMR lines in a salt-containing water and calcium
bicarbonate solution was observed after magnetic field action. This indicates
that the degree of ion hydration is decreased by magnetic treatment. On the
other hand, the width of the resonance line in \textit{distilled water
}remains unchanged after 30 minute treatment in the field $(135\,\,kA/m)$ at
water flow rate of $60\,cm/s\;($Klassen, 1982).

The hydration of diamagnetic ions $(Li^{+},\,\,Mg^{2+},\,\,Ca^{2+})$
decreases, while the hydration of paramagnetic ions $(Fe^{3+},\,\,Ni^{2+}%
,\,\,Cu^{2+})$ increases. It leads from corresponding changes in ultrasound
velocity in ion solutions (Duhanin and Kluchnikov, 1975).

There are numerous data which pointing to an increase the coagulation of
different particles and their sedimentation velocity after magnetic field
treatment. These phenomena provide a reducing the scale formation in heating
systems, widely used in practice. Crystallization and polymerization also
increase in magnetic field. It points to decrease of water activity.

\textbf{Increasing of refraction index (n) and dielectric permeability
$(\epsilon\simeq n^{2})$ and symbatic enhancement of water viscosity (Minenko,
1981) are in total accordance with our viscosity theory (eqs. 11.44 and 11.45
of [1] and article [4]).}

It follows from our mesoscopic model that the increase of (n) is related to
the increase of molecular polarizability $(\alpha)$ due to the shift of
$(a\Leftrightarrow b)_{tr,lb}$ equilibrium of primary effectons leftward under
the action of magnetic field. On the other hand, distant Van der Waals
interactions and consequently dimensions of primary effectons depend on
$\alpha$. This explains the elevation of surface tension of liquids after
magnetic treatment (see Chapter 11 of [1] or [4]).

The leftward shift of $(a\Leftrightarrow b)_{tr,lb}$ equilibrium of primary
effectons must lead to decreasing of water activity due to (n$^{2})$
increasing and structural factor (T/U$_{tot}$) decreasing its structure
ordering. Corresponding changes in the vapor pressure, freezing, and boiling
points, coagulation, polymerization and crystallization are the consequences
of this shift and water activity decreasing.

\smallskip

\textbf{It follows from mesoscopic theory that any changes in condensed matter
properties must be accompanied by change of such parameters as:}

\textbf{1) density;}

\textbf{2) sound velocity;}

\textbf{3) positions of translational and librational bands in oscillatory spectra;}

\textbf{4) refraction index.}

\textbf{\smallskip}

\textbf{Using our equations and computer simulations by means of elaborated
software (CAMP: Comprehensive Analyzer of Matter Properties), it is possible
to obtain from these changes very detailed information (more than 200
parameters) about even small perturbations of matter on meso- and macroscopic levels.}

\smallskip

Available experimental data indicate that all of above mentioned 4
experimental parameters of water have been changed indeed after magnetic treatment.

Minenko (1981) has shown that bidistilled water \textit{density }increases by
about$\,\,0.02\%$ after magnetic treatment $(540\;kA/m$, flow rate $80\;cm/s)
$.

\textit{Sound velocity }in distilled water increases to 0.1\% after treatment
under conditions: $160\;kA/m$ and flow rate $60\;cm/$s.

The \textit{positions of the translational and librational bands }of water
were also changed after magnetic treatment in $415kA/m\;($Klassen, 1982).

\medskip

\begin{center}
{\large Coherent radio-frequency oscillations in water, revealed by C. Smith}

\medskip
\end{center}

\ \textbf{It was shown experimentally by C. Smith (1994) that the water
display a coherent properties. He shows that water is capable of retaining the
frequency of an alternating magnetic field. For a tube of water placed inside
a solenoid coil, the threshold for the alternating magnetic field, potentising
electromagnetic frequencies into water, is 7.6 }$\mu T$\textbf{\ (rms). He
comes to conclusion that the frequency information is carried on\ the magnetic
vector potential.}

\textbf{He revealed also that in a course of yeast cells culture synchronously
dividing, the radio-frequency emission around 1 MHz (10}$^{6}\,1/s)$\textbf{,
7-9 MHz (7-9}$\times10^{6}\,1/s)\;$\textbf{and 50-80 MHz\ (5-9}$\times
$\textbf{10}$^{7}1/s)$\textbf{\ with very narrow bandwidth (\symbol{126}50 Hz)
might be observed for a few minutes.}

\textbf{These frequencies could correspond to frequencies of different water
collective excitations, introduced in our Hierarchic theory, like [lb/tr]
macroconvertons, the [}$a\rightleftharpoons b]_{lb}\,$\textbf{transitons, etc.
(see Fig. 48 of [1] and [2]), taking into account the deviation of water
properties in the colloid and biological systems as respect to pure one.}

\textbf{Cyril Smith has proposed that the increasing of coherence radius in
water could be a consequence of coherent water clusters association due to
Josephson effect (Josephson, 1965): tunneling of molecules between clusters.
As far primary librational effectons are resulted from partial
Bose-condensation of molecules, this idea looks quite acceptable in the
framework of our Hierarchic theory.}

\textbf{The coherent oscillations in tube with water, revealed by C.Smith
could be induced by coherent electromagnetic radiation of microtubules of
cells, produced by correlated intra-MTs water excitations (see Section 17.5
and Fig. 48 of [1]).}

\smallskip

\textbf{The biological effects of magnetically treated water are very
important practically. For example, hemolysis of erythrocytes is more vigorous
in magnetically pretreated physiological solutions (Trincher, 1967). Microwave
radiation induces the same effect (Il'ina et al., 1979). But after boiling
such effects in the treated solutions have been disappeared. It is shown that
magnetic treatment of water strongly stimulates the growth of corn and plants
(Klassen, 1982).}

\textbf{Now it is obvious that a systematic research program is needed to
understand the physical background of multilateral effects of magnetized
water. }

\medskip

\begin{center}
{\large 6. Influence of weak magnetic field on the properties of solid bodies}
\end{center}

{\large \smallskip}

It has been established that as a result of magnetic field action on solids
with interaction energy $(\mu_{B}H)$ much less than kT, many properties of
matter such as hardness, parameters of crystal cells and others change significantly.

The short-time action of magnetic field on silicon semiconductors is followed
by a very long (many days) relaxation process. The action of magnetic field
was in the form of about 10 impulses with a length of 0.2 ms and an amplitude
of about$\;10^{5}A/$m. The most interesting fact was that this relaxation had
an oscillatory character with periods of about several days (Maslovsky and
Postnikov, 1989).

Such a type of long period oscillation effects has been found in magnetic and
nonmagnetic materials.

This points to the general nature of the macroscopic oscillation phenomena in
solids and liquids.

The period of oscillations in solids is much longer than in liquids. This may
be due to stronger deviations of the energy of (\textit{a}) and (\textit{b})
states of primary effectons and polyeffectons from thermal equilibrium and
much lesser probabilities of transiton and deformon excitation. Consequently,
the relaxation time of $(a\Leftrightarrow b)_{tr,lb}$ equilibrium shift in
solids is much longer than in liquids. The oscillations originate due to
instability of dynamic equilibrium between the subsystems of effectons and deformons.

{\large \medskip}

\begin{center}
{\large 7. Possible mechanism of perturbations of nonmagnetic materials }

{\large under magnetic treatment}

\textbf{\smallskip}
\end{center}

\textbf{We shall try to discuss the interaction of magnetic field with
diamagnetic matter like water as an example. The magnetic susceptibility
(}$\chi$\textbf{) of water is a sum of two opposite contributions (Eisenberg
and Kauzmann, 1969):}

\textbf{1) average negative diamagnetic part, induced by external magnetic field:}%

\[
\bar{\chi}^{d}={\frac{1}{2}}(\chi_{xx}+\chi_{yy}+\chi_{zz})\cong
-14.6(\pm1.9)\cdot10^{-6}
\]

\textbf{2) positive paramagnetism related to the polarization of water
molecule due to asymmetry of electron density distribution,\ existing without
external magnetic field. Paramagnetic susceptibility }$(\chi^{p})$\textbf{\ of
}$H_{2}O$\textbf{\ is a tensor with the following components:}%

\begin{equation}
\chi_{xx}^{p}=2.46\cdot10^{-6};\,\,\;\chi_{yy}^{p}=0.77\cdot10^{-6}%
;\,\,\;\chi_{zz}^{p}=1.42\cdot10^{-6}\tag{29}%
\end{equation}
\textbf{The resulting susceptibility:}%

\[
\chi_{H_{2}}=\bar{\chi}^{d}+\bar{\chi}^{p}\cong-13\cdot10^{-6}
\]
\textbf{The second contribution in the magnetic susceptibility of water is
about 10 times lesser than the first one. But the first contribution to the
magnetic moment of water depends on external magnetic field and must disappear
when it is switched out in contrast to second one.}

\medskip

The coherent primary librational effectons of water even in liquid state
contain about 100 molecules $\left[  (n_{M}^{ef})_{lb}\simeq100\right]  $ at
room temperature (Fig. 7a of [1] or Fig.4a of [2]). In ice ($n_{M}^{ef}%
)_{lb}\ge10^{4}$. In (\textit{a})-state the vibrations of all these molecules
are synchronized in the same phase, and in (\textit{b})-state - in
counterphase. Correlation of $H_{2}O$ forming effectons means that the
energies of interaction of water molecules with external magnetic field are additive:%

\begin{equation}
\epsilon^{ef}=n_{M}^{ef}\cdot\mu_{p}H\tag{30}%
\end{equation}
In such a case this total energy of effecton interaction with field may exceed
thermal energy:%

\begin{equation}
\epsilon^{ef}>kT\tag{31}%
\end{equation}
\ \ \ \ 

In the case of polyeffectons formation this inequality becomes much stronger.

It follows from our model that interaction of magnetic field with
(\textit{a})-state of the effectons must be stronger than that with
(\textit{b})-state due to the additivity of the magnetic moments of coherent molecules:%

\begin{equation}
\epsilon_{a}^{ef}>\epsilon_{b}^{ef}\tag{32}%
\end{equation}
Consequently, magnetic field shifts $(a\Leftrightarrow b)_{tr,lb}$ equilibrium
of the effectons leftward. At the same time it minimizes the potential energy
of matter, because potential energy of (\textit{a})-state $(V_{a})$ is lesser
than $(V_{b})$:%

\begin{equation}
V_{a}<V_{b}\text{ \ \ and\thinspace\thinspace\thinspace\ \ }E_{a}%
<E_{b},\tag{33}%
\end{equation}
where $\;E_{a}=V_{a}+T_{\text{kin}}^{a};\;\;E_{b}=V_{b}+T_{\text{kin}}^{b}$
are total energies of the effectons.

\smallskip

\ We keep in mind that the kinetic energies of (\textit{a}) and (\textit{b}%
)-states are equal: $\,\,T_{\text{kin}}^{a}=T_{\text{kin}}^{b}=p^{2}/2m$.

\thinspace\ These energies decreases with increasing of the effectons
dimensions, determined by the most probable impulses in selected directions:%

\[
\lambda_{1,2,3}=h/p_{1,2,3}
\]
The energy of interaction of magnetic field with deformons as a transition
state of effectons must be even less than $\epsilon_{b}^{ef}$ due to lesser
order of molecules in this state and reciprocal compensation of their magnetic moments:%

\begin{equation}
\epsilon_{d}<\epsilon_{b}^{ef}\le\epsilon_{a}^{ef}\tag{34}%
\end{equation}

This important inequality means that as a result of external magnetic field
action the shift of $(a\Leftrightarrow b)_{tr,lb}$ leftward is reinforced by
leftward shift of equilibrium [effectons $\rightleftharpoons$ deformons]
subsystems of matter.

If water is flowing in a tube it increases the relative orientations of all
effectons in volume and stimulate the coherent superclusters formation. All
the above discussed effects must increase. Similar ordering phenomena happen
in a rotating tube with liquid.

After switching off the external magnetic field the relaxation of
\textit{induced ferromagnetism }in water begins. It may be accompanied by the
oscillatory behavior of $(a\Leftrightarrow b)_{tr,lb}$ equilibrium. All the
experimental effects discussed above can be explained as a consequence of
orchestrated in volume $(a\Leftrightarrow b)$ equilibrium oscillations.

\textbf{Remnant ferromagnetism in water was experimentally established using a
SQUID superconducting magnetometer by Kaivarainen et al. in 1992 (unpublished
data). Water was treated in constant magnetic field \thinspace}$50G$%
\textbf{\ \thinspace for two hours. Then it was frozen and after switching off
external magnetic field the remnant ferromagnetism was registered at helium
temperature. Even at this low temperature a slow relaxation time- dependent
decrease of ferromagnetic signal was revealed. These results point to the
correctness of the proposed mechanism of magnetic field - water interaction. }

\textbf{The attempt to make a theory of magnetic field influence on water,
based on other model were made earlier (Yashkichev, 1980). However, this
theory does not take into account the quantum properties of water and cannot
be considered as satisfactory one.\smallskip}

\textbf{The comprehensive material obtained by Udaltsova, Kolombet and Shnol
(1987) when studying various macroscopic oscillations reveals their
fundamental character and their dependence on gravitation factor.}

{\Large The correlated changes of time, entropy and mass of any condensed
matter follows from our theory. }

\bigskip

**************************************************************************

\bigskip

{\Large REFERENCES}

\medskip

\begin{quotation}
\textbf{Babloyantz A. Molecules, Dynamics and Life. An introduction to
self-organization of matter. John Wiley \& Sons, Inc. New York, 1986.}

\textbf{Berezin M.V., Lyapin R.R., Saletsky A.N. Effects of weak magnetic
fields on water solutions light scattering. Preprint of Physical Department of
Moscow University, No.21, 1988. 4 p. (in Russian).}

\textbf{Chernikov F.R. Lightscattering intensity oscillations in water-
protein solutions. Biofizika (USSR}$)\mathbf{\,\,1985,\,31,\;596.}$

\textbf{Chernikov F.R. Effect of some physical factors on light scattering
fluctuations in water and water biopolymer solutions. Biofizika (USSR}%
$)\mathbf{\;}1990a,35,\,711$\textbf{.}

\textbf{Chernikov F.R. Superslow light scattering oscillations in liquids of
different types. Biofizika (USSR}$)\;\mathbf{1990b,\,35,\,\,717.}$

\textbf{Duhanin V.S., Kluchnikov N.G. The problems of theory and practice of
magnetic treatment of water. Novocherkassk, 1975, p.70-73 (in Russian).}

\textbf{Einstein A. Collection of works. Nauka, Moscow, 1965.}

\textbf{Eisenberg D., Kauzmann W. The structure and properties of water.
Oxford University Press, Oxford, 1969.}

\textbf{Egelstaff \thinspace P. A. Static and dynamic structure of liquids and
glasses. J.Non-Crystalline solids.1993, 156, 1-8.}

\textbf{Fild R., Burger M. (Eds.). Oscillations and progressive waves in
chemical systems. Mir, Moscow, 1988. K\"{a}iv\"{a}r\"{a}inen A.I.
Solvent-dependent flexibility of proteins and principles of their function.
D.Reidel Publ.Co., Dordrecht, Boston, Lancaster, 1985,\thinspace pp.290.}

\textbf{K\"{a}iv\"{a}r\"{a}inen A.I. The noncontact interaction between
macromolecules revealed by modified spin-label method. Biofizika
(USSR}$)\;1987,\,32,\,536$\textbf{.}

\textbf{K\"{a}iv\"{a}r\"{a}inen A.I. Thermodynamic analysis of the system:
water-ions-macromolecules. Biofizika (USSR}$),\,\mathbf{1988,\,33,\,549.}$

\textbf{K\"{a}iv\"{a}r\"{a}inen A.I. Theory of condensed state as a
hierarchical system of quasiparticles formed by phonons and three-dimensional
de Broglie waves of molecules. Application of theory to thermodynamics of
water and ice. J.Mol.Liq. }$1989a,\,41,\,53-60$\textbf{.}

\textbf{K\"{a}iv\"{a}r\"{a}inen A.I. Mesoscopic theory of matter and its
interaction with light. Principles of selforganization in ice, water and
biosystems. University of Turku, Finland\thinspace1992, pp.275.}

\textbf{K\"{a}iv\"{a}r\"{a}inen A., Fradkova L., Korpela T. Separate
contributions of large- and small-scale dynamics to the heat capacity of
proteins. A new viscosity approach. Acta Chem.Scand. }%
$\mathbf{1993,47,456-460.}$

\textbf{Kampen N.G., van. Stochastic process in physics and chemistry.
North-Holland, Amsterdam, 1981.}

\textbf{Kiselev V.F., Saletsky A.N., Semikhina L.P. Theor.\thinspace
experim.\thinspace khimya (USSR}$),\mathbf{1988,\,2,\,252-257.}$

\textbf{Klassen V.I. Magnetization of the aqueous systems. Khimiya, Moscow,
1982 (in Russian).}

\textbf{Kozyrev N.A. Causal or nonsymmetrical mechanics in a linear
approximation. Pulkovo. Academy of Science of the USSR. 1958.}

\textbf{Maslovski V.M., Postnikov S.N. In: The treatment by means of the
impulse magnetic field. Proceedings of the IV seminar on nontraditional
technology in mechanical engineering. Sofia-Gorky, 1989.}

\textbf{Minenko V.I. Electromagnetic treatment of water in thermoenergetics.
Harkov, 1981 (in Russian).}

\textbf{Nicolis J.C. Dynamics of hierarchical systems. Springer, Berlin, 1986.}

\textbf{Nicolis J.C., Prigogine I. Self-organization in nonequilibrium
systems. From dissipative structures to order through fluctuations. Wiley and
Sons, N.Y., 1977.}

\textbf{Prigogine I. From Being to Becoming: time and complexity in physical
sciences. W.H.Freeman and Company, San Francisco, 1980.}

\textbf{Prigogine I., Strengers I. \thinspace Order out of chaos.\ Hainemann,
London, 1984.}

\textbf{Semikhina L.P., Kiselev V.F. Izvestiya VUZov. Fizika (USSR), 1988, 5,
13 (in Russian).}

\textbf{Semikhina L.P. Kolloidny jurnal (USSR),\thinspace\ 1981, 43,
\thinspace401.}

\textbf{Shih Y., Alley C.O. Phys Rev.Lett. 1988, 61,\thinspace2921.}

\textbf{Shnol S.E. Physico-chemical factors of evolution. Nauka, Moscow, 1979
(in Russian).}

\textbf{Udaltsova N.B., Kolombet B.A., Shnol S.E. Possible cosmophysical
effects in the processes of different nature. Pushchino, 1987 (in Russian).}

\textbf{Yashkichev V.I. J.Inorganic Chem.(USSR}$),\mathbf{\;}$\textbf{1980,
25, 327.}

\bigskip

\textbf{See also set of previous articles at Los-Alamos archives: \ http://arXiv.org/find/physics/1/au:+Kaivarainen\_A/0/1/0/all/0/1}

\medskip
\end{quotation}
\end{document}